\documentclass[12pt]{article}
\usepackage{epsfig}

\newcommand\pubnumber{IFIC/01-33 \\ FTUV/01-0606}
\newcommand\pubdate{\today}
\newcommand\hepnumber{hep-ph/0106057}

\def\csumb{Departament de F\'{\i}sica Te\`orica, IFIC,
Universitat de Valencia -- CSIC\\
Edifici d'Instituts de Paterna, Apt. Correus 22085, E-46701 Val\`encia,
Spain}
\def\support{\footnote{Work supported by the
European Union TMR network {\it EURODAPHNE}
(Contract No. ERBFMX-CT98-0169) and by DGESIC,
Spain (Grant No. PB97-1261).
}} 

\def\Title#1{\begin{center} {\Large\bf #1 } \end{center}}
\def\Author#1{\begin{center}{ \sc #1} \end{center}}
\def\Address#1{\begin{center}{ \it #1} \end{center}}

\newcommand\pubblock{\rightline{\begin{tabular}{l} \pubnumber\\
         \pubdate\\ \hepnumber \end{tabular}}}
\newenvironment{Abstract}{\begin{quotation}  }{\end{quotation}}
\newenvironment{Presented}{\begin{quotation} \begin{center} 
             Presented at the\end{center}
      \begin{center}\begin{large}}{\end{large}\end{center} \end{quotation}}
\def\Acknowledgments{\bigskip  \bigskip \begin{center}
          \large\bf Acknowledgments\end{center}}

\makeatletter
\def\section{\@startsection{section}{0}{\z@}{5.5ex plus .5ex minus
 1.5ex}{2.3ex plus .2ex}{\large\bf}}
\def\subsection{\@startsection{subsection}{1}{\z@}{3.5ex plus .5ex minus
 1.5ex}{1.3ex plus .2ex}{\normalsize\bf}}
\def\subsubsection{\@startsection{subsubsection}{2}{\z@}{-3.5ex plus
-1ex minus  -.2ex}{2.3ex plus .2ex}{\normalsize\sl}}

\renewcommand{\@makecaption}[2]{%
   \vskip 10pt
   \setbox\@tempboxa\hbox{\small #1: #2}
   \ifdim \wd\@tempboxa >\hsize     
       \small #1: #2\par          
     \else                        
       \hbox to\hsize{\hfil\box\@tempboxa\hfil}
   \fi}

 \def\citenum#1{{\def\@cite##1##2{##1}\cite{#1}}}
 
\newcount\@tempcntc
\def\@citex[#1]#2{\if@filesw\immediate\write\@auxout{\string\citation{#2}}\fi
  \@tempcnta\z@\@tempcntb\m@ne\def\@citea{}\@cite{\@for\@citeb:=#2\do
    {\@ifundefined
       {b@\@citeb}{\@citeo\@tempcntb\m@ne\@citea\def\@citea{,}{\bf ?}\@warning
       {Citation `\@citeb' on page \thepage \space undefined}}%
    {\setbox\z@\hbox{\global\@tempcntc0\csname b@\@citeb\endcsname\relax}%
     \ifnum\@tempcntc=\z@ \@citeo\@tempcntb\m@ne
       \@citea\def\@citea{,}\hbox{\csname b@\@citeb\endcsname}%
     \else
      \advance\@tempcntb\@ne
      \ifnum\@tempcntb=\@tempcntc
      \else\advance\@tempcntb\m@ne\@citeo
      \@tempcnta\@tempcntc\@tempcntb\@tempcntc\fi\fi}}\@citeo}{#1}}
\def\@citeo{\ifnum\@tempcnta>\@tempcntb\else\@citea\def\@citea{,}%
  \ifnum\@tempcnta=\@tempcntb\the\@tempcnta\else
  {\advance\@tempcnta\@ne\ifnum\@tempcnta=\@tempcntb \else\def\@citea{--}\fi
    \advance\@tempcnta\m@ne\the\@tempcnta\@citea\the\@tempcntb}\fi\fi}
\makeatother

\def\beq{\begin{equation}}
\def\eeq#1{\label{#1}\end{equation}}
\def\eeqn{\end{equation}}

\newenvironment{Eqnarray}%
   {\arraycolsep 0.14em\begin{eqnarray}}{\end{eqnarray}}
\def\beqa{\begin{Eqnarray}}
\def\eeqa#1{\label{#1}\end{Eqnarray}}
\def\eeqan{\end{Eqnarray}}

\let\bar=\overbar

\def\etal{{\it et al.}}

\def\Dslash{\not{\hbox{\kern-4pt $D$}}}
\def\dslash{\not{\hbox{\kern-2pt $\del$}}}

\def\msb{{\bar{\ssstyle M \kern -1pt S}}}

\def\lsim{\mathrel{\raise.3ex\hbox{$<$\kern-.75em\lower1ex\hbox{$\sim$}}}}
\def\gsim{\mathrel{\raise.3ex\hbox{$>$\kern-.75em\lower1ex\hbox{$\sim$}}}}

\newcommand{\cO}{{\cal O}}
\newcommand{\cL}{{\cal L}}

\newcommand{\cA}{{\cal A}}
\newcommand{\ba}{\begin{array}{c}}
\newcommand{\bat}{\begin{array}{cc}}
\newcommand{\ea}{\end{array}}
\def\eqn#1{(\ref{#1})}
\newcommand{\no}{\nonumber}
\def\NP{{\sl Nucl. Phys.}}
\def\NPPS{{\sl Nucl. Phys. B (Proc. Suppl.)}}
\def\PL{{\sl Phys. Lett.}}
\def\PRL{{\sl Phys. Rev. Lett.}}
\def\PR{{\sl Phys. Rev.}}

\def\slashchar#1{\setbox0=\hbox{$#1$}\dimen0=\wd0%
\setbox1=\hbox{/}\dimen1=\wd1%
\ifdim\dimen0>\dimen1%
\rlap{\hbox to
\dimen0{\hfil/\hfil}}#1\else                                     
\rlap{\hbox to \dimen1{\hfil$#1$\hfil}}/\fi}
\begin{document}
\begin{titlepage}
\pubblock

\vfill
\def\thefootnote{\fnsymbol{footnote}}
\Title{$\mathbf{\varepsilon'/\varepsilon}$ and Chiral Dynamics}
\vfill
\Author{Antonio Pich\support}
\Address{\csumb}
\vfill
\begin{Abstract}
The long-distance contributions to $K\to 2\pi$ amplitudes
can be pinned down, using well established 
Chiral Perturbation Theory techniques. 
The strong S--wave rescattering of the two
final pions generates sizeable chiral loop corrections, which
have an important impact on the direct CP violation ratio
$\varepsilon'/\varepsilon$ \cite{PP:00,PPS:01}.
Including all large logarithmic corrections, both at short
and long distances,
the Standard Model Prediction for this observable
is found to be \cite{PPS:01}
Re$(\varepsilon'/\varepsilon) = \left( 1.7\pm 0.9\right)\cdot 10^{-3}$,
in good agreement with the most recent experimental measurements.
A better estimate of the strange quark mass could reduce the
theoretical uncertainty to 30\%.
\end{Abstract}
\vfill
\begin{Presented}
5th International Symposium on Radiative Corrections \\ 
(RADCOR--2000) \\[4pt]
Carmel CA, USA, 11--15 September, 2000
\end{Presented}
\vfill
\end{titlepage}
\def\thefootnote{\arabic{footnote}}
\setcounter{footnote}{0}

\section{Introduction}
\label{sec:introduction}

The CP--violating ratio  $\varepsilon'/\varepsilon$  constitutes
a fundamental test for our understanding of flavour--changing
phenomena within the Standard Model framework.
The experimental status has been clarified by the recent
KTEV \cite{KTEV:99}, 
${\rm Re} \left(\varepsilon'/\varepsilon\right) =
(28.0 \pm 4.1) \cdot 10^{-4}$,
and NA48 \cite{NA48:00}, 
${\rm Re} \left(\varepsilon'/\varepsilon\right) =
(15.3 \pm 2.6) \cdot 10^{-4}$,
measurements, which provide clear evidence for a non-zero value and,
therefore, the existence of direct CP violation.
The present world average is \cite{KTEV:99,NA48:00,NA31,E731}, 
\beq\label{eq:exp}
{\rm Re} \left(\varepsilon'/\varepsilon\right) =
(18.0 \pm 2.0) \cdot 10^{-4} \, ,
\qquad\qquad (\chi^2/\mbox{\rm ndf} = 10.8/3).
\eeqn

The theoretical prediction has been rather controversial since
different groups, using different models or approximations,
have obtained different results
\cite{munich,rome,trieste,dortmund,BP:00,NA:00}. 
Although there was no
universal agreement on the $\varepsilon'/\varepsilon$ value
predicted by the Standard Model, it has been often claimed
that it is too small, failing to reproduce the experimental
world average by at least a factor of two.
This claim has generated a very intense theoretical activity,
searching for new sources of CP violation beyond the Standard Model
\cite{beyond}.

It has been pointed out recently \cite{PP:00} that the theoretical 
short--distance evaluations of $\varepsilon'/\varepsilon$ had overlooked
the important role of final--state interactions (FSI) in
$K\to\pi\pi$ decays.
Although it has been known for more than a decade that the
rescattering of the two final pions induces a large correction
to the isospin--zero decay amplitude, this effect was not
taken properly into account in the theoretical predictions.
From the measured $\pi$-$\pi$ phase shifts one can easily infer
\cite{PP:00} 
that FSI generate a strong enhancement of the 
$\varepsilon'/\varepsilon$ prediction,
by roughly the needed factor of two.
This large correction is associated with infrared chiral logarithms
involving the pion mass, which can be rigorously analyzed with
standard Chiral Perturbation Theory ($\chi$PT) techniques
\cite{WE:79,GL:85,EC:95}.
A very detailed analysis, including all large
logarithmic corrections both at short and long distances, has been
presented in ref.~\cite{PPS:01}. The resulting Standard Model
prediction \cite{PPS:01},
\beq\label{eq:th}
{\rm Re} \left(\varepsilon'/\varepsilon\right) =
(17 \pm 9) \cdot 10^{-4} \, ,
\eeqn
is in good agreement with the most recent measurements.

The following sections present a brief overview of the most important
ingredients entering the theoretical prediction of 
$\varepsilon'/\varepsilon\,$:
\begin{enumerate}

\item A short--distance calculation at the electroweak scale and
its renormalization--group evolution to the three--flavour theory
($\mu\lsim m_c$), which sums the leading and
next-to-leading ultraviolet logarithms.

\item The matching to the $\chi$PT description, which so far has
been done at leading order in the $1/N_C$ expansion.

\item Chiral loop corrections, which induce large infrared logarithms
related to FSI.

\end{enumerate}

\section{Theoretical framework}
\label{sec:theory}

In terms of the $K\to\pi\pi$ isospin amplitudes,
$\cA_I = A_I \, e^{\delta_I}$ ($I=0,2$),
\beq
{\varepsilon^\prime\over\varepsilon} =
\; e^{i\Phi}\; {\omega\over \sqrt{2}\vert\varepsilon\vert}\;\left[
{\mbox{Im}(A_2)\over\mbox{Re}(A_2)} - {\mbox{Im}(A_0)\over\mbox{Re}(A_0)}
 \right] \, .
\eeqn
Owing to the well-known ``$\Delta I=1/2$ rule'', 
$\varepsilon'/\varepsilon$ is
suppressed by the ratio
$\omega = \mbox{Re} (A_2)/\mbox{Re} (A_0) \approx 1/22$.
The strong S--wave rescattering of the two final pions generates a
large phase-shift difference between the two isospin amplitudes,
making the phases of $\varepsilon'$ and $\varepsilon$ nearly equal. 
Thus,
\beq
\Phi \approx \delta_2-\delta_0+\frac{\pi}{4}\approx 0 \, .
\eeqn
The CP--conserving amplitudes $\mbox{Re} (A_I)$, their ratio
$\omega$ and $\varepsilon$ are usually set to their experimentally
determined values. A theoretical calculation is then only needed
for the quantities $\mbox{Im} (A_I)$.

One starts above the electroweak scale where the flavour--changing
process, in terms of quarks, leptons and gauge bosons, can be analyzed
within the usual gauge--coupling perturbative expansion
in a rather straightforward way.
Since $M_Z$ is much larger than the long--distance
hadronic scale $M_K$, there are large short--distance logarithmic
contributions which can be summed up using the
Operator Product Expansion (OPE) \cite{WI:69} and the renormalization
group. The proper way to proceed makes use of modern
Effective Field Theory (EFT) techniques \cite{EFT}.

The renormalization group is used to evolve down in energy 
from the electroweak scale, where the top quark and the $Z$ and 
$W^\pm$ bosons are integrated out. That means that one changes to a
different EFT where those heavy particles are no longer 
explicit degrees of freedom. The new Lagrangian contains a tower of
operators constructed with the light fields only, which scale as
powers of $1/M_Z$. The information on the heavy fields is hidden in
their (Wilson) coefficients, which are fixed by
``matching'' the high-- and low--energy theories at the point $\mu=M_Z$.
One follows the evolution further to lower energies, using the
EFT renormalization group equations, until a new particle
threshold is encountered. Then, the whole procedure of integrating the
new heavy scale and matching to another EFT starts again.

{\renewcommand{\arraystretch}{1.0}
\begin{figure}[tbh]    
\setlength{\unitlength}{0.75mm} \centering
\begin{picture}(165,120)
\put(0,0){\makebox(165,120){}}
\thicklines
\put(10,105){\makebox(40,15){\large Energy Scale}}
\put(58,105){\makebox(36,15){\large Fields}}
\put(110,105){\makebox(40,15){\large Effective Theory}}
\put(8,108){\line(1,0){149}} {\large
\put(10,75){\makebox(40,27){$M_Z$}}
\put(58,75){\framebox(36,27){$\ba W, Z, \gamma, g \\
     \tau, \mu, e, \nu_i \\ t, b, c, s, d, u \ea $}}
\put(110,75){\makebox(40,27){Standard Model}}

\put(10,40){\makebox(40,18){$\lsim m_c$}}
\put(58,40){\framebox(36,18){$\ba  \gamma, g  \; ;\; \mu ,  e, \nu_i
             \\ s, d, u \ea $}}
\put(110,40){\makebox(40,18){$\cL_{\mathrm{QCD}}^{(n_f=3)}$, \
             $\cL_{\mathrm{eff}}^{\Delta S=1,2}$}}

\put(10,5){\makebox(40,18){$M_K$}}
\put(58,5){\framebox(36,18){$\ba\gamma \; ;\; \mu , e, \nu_i  \\
            \pi, K,\eta  \ea $}}
\put(110,5){\makebox(40,18){$\chi$PT}}
\linethickness{0.3mm}
\put(76,37){\vector(0,-1){11}}
\put(76,72){\vector(0,-1){11}}
\put(80,64.5){OPE}
\put(80,29.5){$N_C\to\infty$}}
\end{picture}
\caption[0]{Evolution from $M_Z$ to $M_K$.
  \label{fig:eff_th}}
\end{figure}
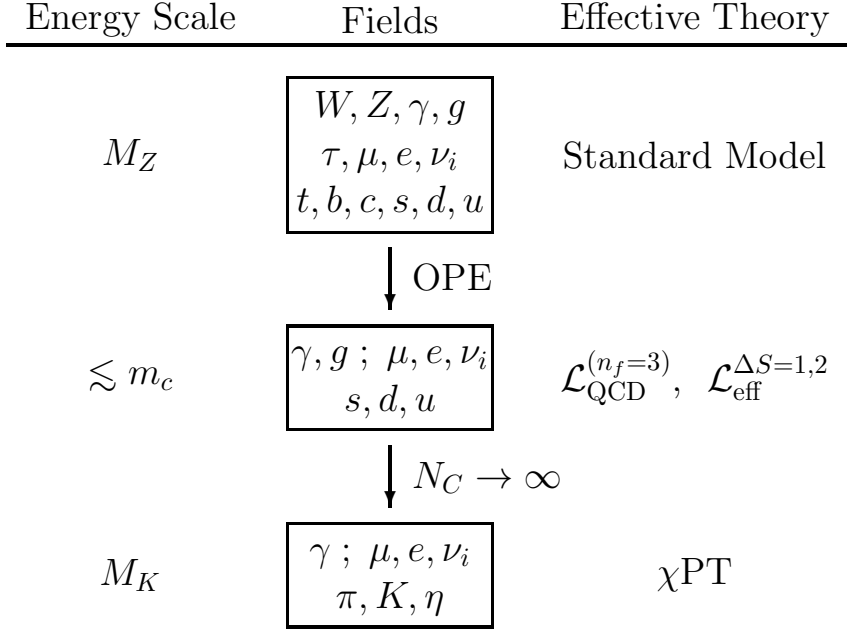
}

One proceeds down to scales $\mu < m_c$ and
gets finally an effective $\Delta S=1$ Lagrangian, defined in the
three--flavour theory \cite{GW:79,BU:99},
\beq  
 {\cal L}_{\mathrm eff}^{\Delta S=1}= - \frac{G_F}{\sqrt{2}}
 V_{ud}^{\phantom{*}}\,V^*_{us}\,  \sum_{i=1}^{10}
 C_i(\mu) \; Q_i (\mu) \; ,
 \label{eq:lag}
\eeq{eq:Leff}
which is a sum of local four--fermion operators $Q_i$,
constructed with the light degrees of freedom, modulated
by Wilson coefficients $C_i(\mu)$ which are functions of the
heavy masses.
We have explicitly factored out the Fermi coupling $G_F$
and the Cabibbo--Kobayashi--Maskawa (CKM) matrix elements
$V_{ij}$ containing the usual Cabibbo suppression of $K$ decays.
The unitarity of the CKM matrix allows to write
%
\beq
C_i(\mu) =  z_i(\mu) + \tau\ y_i(\mu) \; , 
\label{eq:Lqcoef} 
\eeqn
where 
$\tau = - V_{td}^{\phantom{*}} V_{ts}^{*}/V_{ud}^{\phantom{*}} V_{us}^{*}$.  
Only the $y_i$ components are needed to determine the CP--violating 
decay amplitudes.
The overall renormalization scale $\mu$ separates
the short-- ($M>\mu$) and long-- ($m<\mu$) distance contributions,
which are contained in $C_i(\mu)$ and $Q_i$, respectively.
The physical amplitudes are of course independent of $\mu$.

Our knowledge of $\Delta S=1$ transitions has improved qualitatively
in recent years, thanks to the completion of the next-to-leading
logarithmic order calculation of the Wilson coefficients
\cite{buras1,ciuc1}.
All gluonic corrections of $O(\alpha_s^n t^n)$ and
$O(\alpha_s^{n+1} t^n)$ are known,
where $t\equiv\ln{(M_1/M_2)}$ refers to the logarithm of any ratio of
heavy mass scales $M_1,M_2\geq\mu$.
Moreover, the full $m_t/M_W$ dependence (at lowest
order in $\alpha_s$) is taken into account.

In order to predict physical amplitudes, however, one is still
confronted with the calculation of hadronic matrix elements of
the four--quark operators. This is a very difficult problem,
which so far remains unsolved.
Those matrix elements are usually parameterized
in terms of the so-called bag parameters $B_i$, which measure them
in units of their vacuum insertion approximation values.

To a very good approximation, the Standard Model prediction for
$\varepsilon'/\varepsilon$ can be written (up to global factors)
as \cite{munich}
\beq
{\varepsilon'\over\varepsilon} \sim
\left [ B_6^{(1/2)}(1-\Omega_{IB}) - 0.4 \, B_8^{(3/2)}
 \right ]\, .
\label{EPSNUM}
\eeqn
Thus, only two operators are numerically relevant:
the QCD penguin operator $Q_6$ ($\Delta I=1/2$)
governs $\mbox{Im}(A_0)$, while $\mbox{Im}(A_2)$ ($\Delta I=3/2$)
is dominated by the electroweak penguin operator $Q_8$.
The parameter
\beq
\Omega_{IB} = {1\over \omega}
{\mbox{Im}(A_2)_{IB}
\over \mbox{Im}(A_0)}
\label{eq:isospin}
\eeqn
takes into account isospin--breaking corrections, which get enhanced
by the large factor $1/\omega$.

The isospin--breaking correction coming from $\pi^0$-$\eta$  mixing
was originally estimated to be $\Omega_{IB}^{\pi^0\eta}=0.25$ 
\cite{Omega,BG:87}. Together with the usual ansatz $B_i\sim 1$, 
this produces a large numerical cancellation
in eq.~\eqn{EPSNUM} leading to low values of 
$\varepsilon'/\varepsilon$ around $7\cdot 10^{-4}$.
A recent improved calculation of $\pi^0$-$\eta$  mixing
at $\cO(p^4)$ in $\chi$PT has found the result \cite{EMNP:00}
\beq
\Omega_{IB}^{\pi^0\eta}\, =\, 0.16\pm 0.03 \, .
\label{eq:omIB}
\eeqn
This smaller number, slightly increases the naive estimate of 
$\varepsilon'/\varepsilon$.

\section{Chiral Perturbation Theory}
\label{sec:ChPT}

Below the resonance region
one can use global symmetry considerations to define another
EFT in terms of the QCD Goldstone bosons
($\pi$, $K$, $\eta$). The $\chi$PT formulation of the Standard Model
\cite{WE:79,GL:85,EC:95} describes
the pseudoscalar--octet dynamics, through a perturbative expansion 
in powers of momenta and quark masses
over the chiral symmetry breaking scale
($\Lambda_\chi\sim 1\; {\rm GeV}$).

Chiral symmetry fixes the allowed $\chi$PT operators.
At lowest order in the chiral expansion,
the most general effective bosonic Lagrangian
with the same $SU(3)_L\otimes SU(3)_R$ transformation properties
as the short--distance Lagrangian \eqn{eq:Leff} contains three terms,
transforming as $(8_L,1_R)$, $(27_L,1_R)$ and $(8_L,8_R)$, respectively.
Their corresponding chiral couplings are denoted by
$g_8$, $g_{27}$ and $g_{ew}$.

The tree--level $K\to\pi\pi$ amplitudes generated
by the lowest--order $\chi$PT Lagrangian do not contain any strong phases:
\beqa
\cA_0 & =& -{G_F\over \sqrt{2}} \, V_{ud}^{\phantom{*}} V_{us}^*
\;\sqrt{2}\,
f_\pi\;\left\{\left ( g_8+{1\over 9}\, g_{27}\right )(M_K^2-M_\pi^2)
-{2\over 3}\, f_\pi^2\, e^2 \; g_8\; g_{ew}\right\}\; ,
\nonumber\\
\cA_2& =&  -{G_F\over \sqrt{2}} \, V_{ud}^{\phantom{*}} V_{us}^*
\; {2\over 9}
\,f_\pi\;\biggl\{ 5\, g_{27}\, (M_K^2-M_\pi^2) -3\, f_\pi^2\, e^2\; 
g_8\; g_{ew}\biggr\}
\; .
\label{TREE}
\eeqan
Taking the measured phase shifts into account,
the moduli of $g_8$ and $g_{27}$ can be extracted from
the CP--conserving $K \rightarrow 2 \pi$ decay rates;
one gets \cite{PGR:86}
$|g_8|\approx 5.1$ and $|g_{27}|\approx 0.29$. 
The huge difference between these two couplings
shows the well--known
enhancement of octet $\vert\Delta I\vert = 1/2$ transitions.
The $g_{ew}$ term
is the low--energy realization of the electroweak penguin operator.

The isospin amplitudes $\cA_I$ have been
computed up to next--to--leading order in the chiral expansion
\cite{PPS:01,KA91,BPP,EIMNP:00,CDG:99,EKW:93}.
The only remaining problem is the calculation of the $\chi$PT couplings
from the effective short--distance Lagrangian \eqn{eq:Leff},
which requires to perform the matching between the two EFTs.
This can be easily done in the large--$N_C$ limit of QCD
\cite{HO:74}, because
in this limit the four--quark operators factorize into currents
which have well--known chiral realizations.
The local $\cO(p^4)$ contributions to the amplitudes $\cA_I$
can be easily included in eqs.~\eqn{TREE},
through effective correction factors $\Delta_C\cA_I^{(R)}$ to the
lowest--order $g_R$ contributions.
At leading order in $1/N_C$, one gets \cite{PPS:01}:
\beqa\label{eq:NC_results}
\lefteqn{
g_8^\infty\,\left[ 1 + \Delta_C\cA_0^{(8)} \right]^\infty\, =} &&
\nonumber\\ &&
\left\{ 
 -{2\over 5}\,C_1(\mu)+{3\over 5}\,C_2(\mu)+C_4(\mu)
- 16\, L_5\, C_6(\mu)\,
\left[ {M_K^2 \over (m_s + m_q)(\mu)\, f_\pi}\right]^2\right\}\,
f_0^{K\pi}(M_\pi^2)   \, ,
\nonumber\\  &&\nonumber\\
\lefteqn{
g_{27}^\infty\,\left[ 1 + \Delta_C\cA_0^{(27)}\right]^\infty \ = \
g_{27}^\infty\,\left[ 1 + \Delta_C\cA_2^{(27)}\right]^\infty \, = \,
{3\over 5}\,[C_1(\mu)+C_2(\mu)]\;  f_0^{K\pi}(M_\pi^2) \, , }&&
\nonumber\\ &&\nonumber\\
\lefteqn{
e^2\, g_8^\infty\, \left[ g_{ew} + \Delta_C\cA_0^{(ew)}\right]^\infty 
\, =\,
-3\, C_8(\mu)\,\left[ {M_K^2 \over (m_s + m_q)(\mu)\, f_\pi}\right]^2
\left[1 + {4 L_5\over f_\pi^2} M_K^2\right]
} &&
\nonumber\\ &&\hspace{4.3cm}\mbox{}
- {3\over 4} \left[ C_7- C_9 + C_{10}\right]\!(\mu) \:
{M_K^2-M_\pi^2\over f_\pi^2}\; f_0^{K\pi}(M_\pi^2)\, ,
\nonumber\\ &&\nonumber\\
\lefteqn{
e^2\, g_8^\infty\, \left[ g_{ew} + \Delta_C\cA_2^{(ew)}\right]^\infty
 \, = \,
-3\, C_8(\mu)\,\left[ {M_K^2 \over (m_s + m_q)(\mu)\, f_\pi}\right]^2
\left[1 + {4 L_5\over f_\pi^2} M_\pi^2\right]
}&&\hfill\mbox{}
\nonumber\\ &&\hspace{4.3cm}\mbox{}
+ {3\over 2} \left[ C_7 - C_9 - C_{10}\right]\!(\mu)\:
{M_K^2-M_\pi^2\over f_\pi^2}\; f_0^{K\pi}(M_\pi^2) \, ,
\eeqan
where 
$f_0^{K\pi}(M_\pi^2)\approx 1 + 4 L_5\, M_\pi^2/f_\pi^2 $
is the $K\pi$ scalar form factor at the pion mass scale,
$L_5$ is a coupling of the $\cO(p^4)$ strong chiral Lagrangian
and $m_q\equiv m_u = m_d$.
In the limit $N_C\to\infty$,
$L_5^\infty = {1\over 4} f_\pi^2 
\left( {f_K\over f_\pi} - 1 \right) /
\left(M_K^2-M_\pi^2\right) \approx 2.1\cdot 10^{-3}$
and \ $f_0^{K\pi}(M_\pi^2)\approx 1.02$.

These results are equivalent to the
usual large--$N_C$ evaluations of the $B_i$ factors.
In particular, for $\varepsilon'/\varepsilon$ 
where only the imaginary part of
the $g_R$ couplings matter [i.e. Im($C_i$)] they
amount to $B_8^{(3/2)}\approx B_6^{(1/2)}=1$. Therefore, up to minor
variations on some input parameters, the corresponding 
$\varepsilon'/\varepsilon$ prediction 
reproduces the published results of the Munich
\cite{munich} and Rome \cite{rome} groups.

The large--$N_C$ limit is only applied to the matching between
the 3--flavour quark theory and $\chi$PT,
as indicated in Figure~\ref{fig:eff_th}.
The evolution from the electroweak
scale down to $\mu < m_c$ has to be done without any unnecessary expansion
in powers of $1/N_C$; otherwise, one would miss large corrections
of the form ${1\over N_C} \ln{(M/m)}$, with $M\gg m$ two widely
separated scales \cite{BBG87}.
Thus, the Wilson coefficients contain the full $\mu$ dependence.

The large--$N_C$ factorization of the four--quark operators $Q_i$
($i\not=6,8$) does not provide any scale dependence.
Since the anomalous
dimensions of these operators vanish when $N_C\to\infty$ \cite{BBG87},
a very important ingredient is lost in this limit \cite{PI:89}.
To achieve a reliable expansion in powers of $1/N_C$,
one needs to go to the next order where this physics is captured
\cite{PI:89,PR:91}. This is the reason why the study of the $\Delta I=1/2$
rule has proven to be so difficult. Fortunately, these operators
are numerically suppressed in the 
$\varepsilon'/\varepsilon$ prediction
and their contributions can be easily corrected with the
information provided by the measured CP--conserving rates
\cite{PPS:01,munich}.

The only anomalous dimensions which survive when $N_C\to\infty$
are precisely the ones corresponding to $Q_6$ and $Q_8$
\cite{BG:87,BBG87}. One can then expect that the matrix elements
of these two operators
are well approximated by this limit \cite{PI:89,PR:91,JP:94}.
These operators  factorize into colour--singlet
scalar and pseudoscalar currents, which are $\mu$ dependent.
This generates the factors
$\langle\bar q q\rangle(\mu) 
\approx - {f_\pi^2 M_K^2/ (m_s + m_q)(\mu)}$,
which exactly cancel the $\mu$ dependence of
$C_{6,8}(\mu)$ at large $N_C$ \cite{BG:87,BBG87,PI:89,PR:91,JP:94,dR:89}.
It remains of course a dependence at next-to-leading order.

Therefore, while there are large $1/N_C$ corrections to Re($g_I$)
\cite{PR:91}, the large--$N_C$ limit is expected to give a
good estimate of Im($g_I$).

\section{Chiral loop corrections}
\label{sec:loops}

The lowest--order calculation does not provide any strong phases
$\delta_I$. Those phases originate in the
final rescattering of the two pions and, therefore, are generated by
chiral loops which are of higher order in both the momentum
and $1/N_C$ expansions.
Analyticity and unitarity require the presence of a corresponding
dispersive FSI effect in the moduli of the isospin amplitudes.
Since the S--wave strong phases are quite large,
specially in the isospin--zero case,
one should expect large higher--order unitarity corrections.

The size of the FSI effect can be calculated at one loop in $\chi$PT.
The dominant one-loop correction to the octet amplitude
comes indeed from the
elastic soft rescattering of the two pions in the final state.
The existing one--loop analyses \cite{PPS:01,KA91,BPP}
show that pion
loop diagrams provide an important enhancement of the $\cA_0$ amplitude,
implying a corresponding reduction of the phenomenologically fitted 
value of $|g_8|$.
This chiral loop correction destroys the accidental numerical
cancellation in eq.~\eqn{EPSNUM}, generating a sizeable enhancement
of the $\varepsilon'/\varepsilon$ prediction \cite{PP:00}.

Let us decompose the isospin amplitudes in their different
chiral components as
${\cal A}_0 = {\cal A}_0^{(8)} + {\cal A}_0^{(27)}
+ {\cal A}_0^{(ew)}$ 
and
${\cal A}_2 = {\cal A}_2^{(27)} + {\cal A}_2^{(ew)}$.
Moreover, we can write them in the form
\beq
{\cal A}_I^{(R)} \; = \; {\cal A}_I^{(R)\infty} \;\times\;\:
{\cal C}_I^{(R)} \, ,
\eeqn
where ${\cal A}_I^{(R)\infty}$ are the large--$N_C$ results
obtained in the previous section.
The correction factors 
${\cal C}_I^{(R)}\equiv 1 + \Delta_L{\cal A}_I^{(R)}$ 
contain the chiral loop contributions $\Delta_L{\cal A}_I^{(R)}$
that we are interested in.
Their complete analytical expressions at one loop
in $\chi$PT
have been given in ref.~\cite{PPS:01}, where
the following numerical values have been obtained:
\beqa
{\cal C}_0^{(8)}&\; =\; & 
1.27 \pm 0.05 +  0.46 \ i \, ,\no  \\
{\cal C}_0^{(27)}&= & 
 2.0 \pm 0.7 + 0.46 \ i \, ,\no  \\
{\cal C}_0^{(ew)}&= & 
 1.27  \pm 0.05 + 0.46 \ i \, ,\no  \\
{\cal C}_2^{(27)}&= & 
 0.96 \pm 0.05-0.20 \ i\, ,\no  \\
{\cal C}_2^{(ew)}&= & 
 0.50  \pm 0.24-0.20 \ i \, .
\label{eq:onel}
\eeqan
The central values have been evaluated at the chiral renormalization
scale $\nu = M_\rho$. To estimate the corresponding uncertainties
we have allowed the scale $\nu$ to change between 0.6 and 1~GeV.
The scale dependence is only present in the dispersive contributions
and should cancel with the corresponding $\nu$
dependence of the local $\chi$PT counterterms. However, this dependence is
next-to-leading in $1/N_C$ and, therefore, is not included in the
large--$N_C$ determination of the chiral couplings.
The sensitivity of the results to the scale $\nu$
gives a good estimate of those missing contributions.
Notice that all amplitudes with a given isospin get the same
absorptive contribution, as it should since they have identical strong
phase shifts.

The numerical corrections to the 27--plet amplitudes do not have 
much phe\-no\-menological interest for CP--violating observables, 
because ${\rm Im}(g_{27}) = 0$.
Remember that the CP--conserving amplitudes ${\rm Re}(A_I)$
are set to their experimentally determined values.
What is relevant for the $\varepsilon'/\varepsilon$
prediction is the 35\% enhancement of the isoscalar octet
amplitude \ Im[$A_0^{(8)}$] \ and the 46\% reduction of \
Im[$A_2^{(ew)}$]. Just looking to the simplified formula
(\ref{EPSNUM}), one realizes immediately the obvious impact of
these one-loop chiral corrections. 

A complete $\cO(p^4)$ calculation \cite{EMNP:00,EIMNP:00}
of the isospin--breaking parameter $\Omega_{IB}$ is  not yet
available. The value 0.16 quoted in eq.~(\ref{eq:omIB}) 
only accounts for the contribution from $\pi^0$--$\eta$ 
mixing \cite{EMNP:00}
and should be corrected by the effect of chiral loops. 
Since $|{\cal C}_2^{(27)}| \approx 0.98\pm 0.05$,
one does not expect any large correction of \ ${\rm Im}(A_2)_{IB}$,
while we know that
Im[$A_0^{(8)}$] gets enhanced by a factor 1.35.
Taking this into account, one gets the corrected value
\beq
\Omega_{IB} \;\approx\;\Omega_{IB}^{\pi^0\eta}\;
\left|{{\cal C}_2^{(27)}\over {\cal C}_0^{(8)}}\right|
\; = \; 0.12 \pm 0.05 \, ,
\eeqn
where the quoted error is an educated theoretical guess.
This value agrees with the result 
$\Omega_{IB} = 0.08 \pm 0.05\pm 0.01$,
obtained in ref.~\cite{MW:00} by using three different
models \cite{trieste,EKW:93,PR:91,EGPR:89,EPR:91,GV:99}
to estimate the relevant $\cO(p^4)$ chiral couplings.

The sensitivity to higher--order
chiral loop corrections has been investigated in
ref.~\cite{PPS:01} through
an Omn\`es exponentiation of the dominant pion loops
\cite{PP:00},
using the experimental $\pi\pi$ phase shifts.
The standard one-loop $\chi$PT results and the Omn\`es
calculation agree within errors, indicating a good
convergence of the chiral expansion.

\section{Numerical results and discussion}
\label{sec:numerics}

The infrared effect of chiral loops generates an important enhancement
of the isoscalar $K\to\pi\pi$ amplitude. This effect gets amplified
in the prediction of $\varepsilon'/\varepsilon$, because 
at lowest order (in both $1/N_C$ and the chiral expansion) there
is an accidental numerical cancellation between the $I=0$ and $I=2$
contributions. Since the chiral loop corrections destroy this cancellation,
the final result for $\varepsilon'/\varepsilon$ is dominated by the
isoscalar amplitude. Thus, the Standard Model prediction for
$\varepsilon'/\varepsilon$ is finally governed by the matrix element
of the gluonic penguin operator $Q_6$.

A detailed numerical analysis has been provided in
ref.~\cite{PPS:01}.
The short--distance Wilson coefficients have been evaluated
at the scale $\mu = 1$ GeV. Their associated uncertainties
have been estimated through the sensitivity to 
changes of $\mu$ in the range $M_\rho < \mu < m_c$
and to the choice of $\gamma_5$ scheme.
Since the most important $\alpha_s$ corrections appear at the 
low--energy scale $\mu$, the strong coupling
has been fixed  at the $\tau$ mass,
where it is known~\cite{pich1} with about a few percent level of accuracy:
$\alpha_s(m_\tau)=0.345\pm 0.020$.
The values of $\alpha_s$ at the other needed scales can be
deduced through the standard renormalization group evolution.

Taking the experimental value of $\varepsilon$,
the CP--violating ratio $\varepsilon'/\varepsilon$ 
is proportional to the CKM factor
Im$(V^*_{ts} V^{\phantom{*}}_{td}) = (1.2\pm 0.2) \cdot 10^{-4}$
\cite{UT:00}. 
This number is sensitive to the input values of
several non-perturbative hadronic parameters adopted 
in the usual unitarity triangle analysis; thus, it is subject to
large theoretical uncertainties which are difficult to quantify
\cite{PP:95}.
Using instead the theoretical prediction of $\varepsilon$, this CKM
factor drops out from the ratio $\varepsilon'/\varepsilon$; 
the sensitivity
to hadronic inputs is then reduced to the explicit remaining dependence
on the $\Delta S=2$ scale--invariant bag parameter $\hat{B}_K$.
In the large--$N_C$ limit, $\hat{B}_K = 3/4$.
We have performed the two types of numerical analysis, obtaining
consistent results. This allows us to estimate better the theoretical
uncertainties, since the two analyses have different sensitivity
to hadronic inputs.

The final result quoted in ref.~\cite{PPS:01} is:
\beq
\mbox{\rm Re}\left(\varepsilon'/\varepsilon\right) \; =\;  
\left(1.7\pm 0.2\, {}_{-0.5}^{+0.8} \pm 0.5\right) \cdot 10^{-3}
\; =\; \left(1.7\pm 0.9\right) \cdot 10^{-3}\, .
\label{eq:final_result}
\eeq{eq:SMpred}
The first error comes from the short--distance evaluation
of Wilson coefficients and the choice of low--energy
matching scale $\mu$.
The uncertainty coming from varying  the strange quark mass 
in the interval $(m_s+ m_q)(1\, \rm{GeV})=156\pm 25\, \rm{MeV}$
\cite{ms2000,PP:99,ALEPH:99,KM:00,JA:98,LatMass,BPR:98} is indicated 
by the second error.
The most critical step is the matching between the short-- and 
long--distance descriptions. 
We have performed this matching at leading
order in the $1/N_C$ expansion, where the result is known
to $\cO(p^4)$ and $\cO(e^2p^2)$ in $\chi$PT.
This can be expected to provide a good approximation to the matrix 
elements of the leading $Q_6$ and $Q_8$ operators. Since all ultraviolet
and infrared logarithms have been resummed, our educated guess for
the theoretical uncertainty associated with $1/N_C$ corrections
is $\sim 30\%$ (third error).

Thus, a better determination of the strange quark mass would allow
to reduce the uncertainty to the 30\% level.
In order to get a more accurate prediction, it would be necessary to have
a good analysis of next--to--leading $1/N_C$ corrections. This is
a very difficult task, but progress in this direction can be
expected in the next few years 
\cite{trieste,BP:00,PR:91,DG:00,NLO_NC,latka}.

To summarize, using a well defined computational scheme,
it has been possible to pin down the value of 
$\varepsilon'/\varepsilon$ with an acceptable accuracy.
Within the present uncertainties, the resulting
Standard Model theoretical prediction \eqn{eq:SMpred}
is in good agreement with the measured experimental
value \eqn{eq:exp}, without any need to invocate a new
physics source of CP violation.

\Acknowledgments
I warmly thank Elisabetta Pallante and Ignazio Scimemi
for a rewarding collaboration,
and the RADCOR-2000 organizers for making possible
a very stimulating workshop in such a nice environment.


\begin{thebibliography}{99}
\bibitem{beyond} There is a vast literature on this issue,
which can be traced back from the recent review by
A. Masiero and O. Vives, hep-ph/0104027.
\end{thebibliography}
\end{document}